# تطوير مدوّنة تدريب لتحسين أداء نظم فصل المتكلمين المعتمدة على الذكاء الصنعي

م. رواد ملحم[1]، د.م. آصف جعفر[2]، د.م. أميمة الدكاك[3]

[1]طالب دكتوراه في المعهد العالي للعلوم التطبيقية والتكنولوجيا، معالجة إشارة، ذكاء صنعي.

[2]دكتور أستاذ في المعهد العالي للعلوم التطبيقية والتكنولوجيا، ذكاء صنعي، معالجة إشارة.

[3]دكتور أستاذ في المعهد العالي للعلوم التطبيقية والتكنولوجيا، معالجة إشارة كلامية، ذكاء صنعي.

## الملخص

تعتبر مشكلة فصل المتكلمين مسألة بحث مفتوحة وتحتاج إلى الكثير من العمل بالرغم من النتائج المنافسة التي ظهرت في السنوات الأخيرة، حيث تنخفض تلك النتائج كثيراً عند فصل المتكلمين في ظروف تسجيل حقيقية (ضجيج– صدى– تداخلات). يرجع السبب في ذلك التدهور إلى تدريب النماذج العصبونية على مدونات تدريب تركيبية تتكون من إشارات صوتية هي إشارة مزيج لصوتين والأصوات المفردة Ground Truths التي شكلت ذلك المزيج. صُممت إشارات المزيج في المدونات التركيبية باستخدام برامج محاكاة حاسوبية، لا تعكس بشكل كاف إشارات المزيج الواقعية التي يلتقطها الميكرفون. لا يوجد حتى الآن مدوّنة تدريب حقيقية أو واقعية لفصل المتكلمين، والعائق الرئيسي في ذلك هو صعوبة الحصول على الأصوات المفردة بعد تسجيل إشارة المزيج. نقدّم في هذه الورقة طريقة لبناء أول مدوّنة تدريب حقيقية لفصل المتكلمين تتضمن إشارات المزيج مع الأصوات المفردة الموافقة لكل مزيج. اختبرنا هذه المدوّنة على نموذج تعلم عميق وقارناه مع مدوّنة تركيبية حيث لاحظنا تحسّن دقة فصل المتكلمين بمقدار 1.65 dB حسب المعيار Scale Invariant Signal to Distortion Ratio (SI-SDR) في حالة المزج الحقيقي. أظهرت النتائج أهمية مجموعات التدريب الحقيقية في تحسين أداء خوارزميات فصل المتكلمين في بيئات حقيقية.

**الكلمات المفتاحية:** فصل المتكلمين، مدوّنة التدريب، الأصوات المفردة، إشارة مزيج.

# Developing an Effective Training Dataset to Enhance the Performance of AI-based Speaker Separation Systems

## Rawad Melhem[1], Assef Jafar[2], Oumayma Al Dakkak[3]

[1] PhD student at Higher Institute for Applied Sciences and Technology, signal processing, Artificial Intelligence.
rawad.melhem@hiast.edu.sy
[2] Professor at Higher Institute for Applied Sciences and Technology, Signal Processing, Artificial Intelligence.
assef.jafar@hiast.edu.sy
[3] Professor at Higher Institute for Applied Sciences and Technology, Speech Signal Processing, Artificial Intelligence.
oumayma.dakkak@hiast.edu.sy

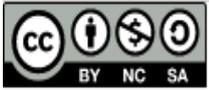

## Abstract

This paper addresses the challenge of speaker separation, which remains an active research topic despite the promising results achieved in recent years. These results, however, often degrade in real recording conditions due to the presence of noise, echo, and other interferences. This is because neural models are typically trained on synthetic datasets consisting of mixed audio signals and their corresponding ground truths, which are generated using computer software and do not fully represent the complexities of real-world recording scenarios. The lack of realistic training sets for speaker separation remains a major hurdle, as obtaining individual sounds from mixed audio signals is a non-trivial task. To address this issue, we propose a novel method for constructing a realistic training set that includes mixture signals and corresponding ground truths for each speaker. We evaluate this dataset on a deep learning model and compare it to a synthetic dataset. We got a 1.65 dB improvement in Scale Invariant Signal to Distortion Ratio (SI-SDR) for speaker separation accuracy in realistic mixing. Our findings highlight the potential of realistic training sets for enhancing the performance of speaker separation models in real-world scenarios.

# 1- مقدمة

طرح الباحث Cherry عام 1953 تساؤلاً عن قدرة الانسان الانتقائية في تمييز صوت ضمن مزيج من الأصوات [1] معرّفاً بذلك محور بحث لم يُغلق حتى يومنا هذا على الرغم من مرور أكثر من نصف قرن على طرحه. أُطلق على المسألة حفلة الكوكتيل Cocktail Party، وتُعرّف على الشكل التالي: يلتقط ميكرفون مجموعة من الأصوات المتداخلة في بيئة ما، والمطلوب استخراج إشارة كل منبع صوتي لوحده. تُعتبر هذه المسألة من أعقد المسائل في البحث العلمي، ويعود ذلك لعدة أسباب أهمها: تنوع الأصوات التي يلتقطها الميكرفون (كلام، موسيقى، ضجيج) فكل نوع له طبيعة مختلفة عن البقية، عدم استقرار الإشارة الكلامية (وهي الإشارة المستهدفة في فصل الأصوات) فطيفها الترددي يتغير باستمرار مع الزمن، عدد المنابع الصوتية غير معروف خلال التسجيل لذلك يطلق على المسألة أيضاً الفصل الأعمى. إضافة إلى ذلك، فإن دقة فصل المنابع تتناسب عكساً مع عدد المنابع الموجودة ضمن إشارة المزيج.

إن مسألة فصل المتكلمين هي حالة خاصة من حفلة الكوكتيل فهي تختص بفصل الإشارات الكلامية بعضها عن بعض، دون معالجة الصدى والضجيج وفصل الموسيقى. يفيد فصل الإشارات الكلامية في الكثير من التطبيقات أهمها:

- تطبيقات تحويل الكلام إلى نص مكتوب Automatic meeting transcription.
- تسمية المقاطع الصوتية بشكل آلي Automatic captioning for audio/video recordings مثل موقع YouTube.
- التطبيقات التي تحتاج لتفاعل الانسان مع الآلة عن طريق الصوت كما في تجهيزات انترنت الأشياء.
- الأجهزة المساعدة للسمع Hearing aids.

نطرح في هذا العمل طريقة هي الأولى حتى الآن لبناء مدوّنة تدريب حقيقية مع الأصوات المفردة، وذلك باستخدام تابع ضمن برنامج MATLAB الذي يتخاطب مع بطاقة الصوت ويستطيع تشغيل وتسجيل الملفات الصوتية بآن واحد وفي الزمن الحقيقي. جرى تنظيم ما تبقى من الورقة كما يلي: الفقرة (2)، تعرض الدراسة المرجعية، تبين الفقرة (3) الفرق بين المدوّنات التركيبية والمدوّنات الحقيقية، تشرح الفقرة (4) الطريقة المقترحة لبناء المدوّنة الحقيقية، تفصّل الفقرة (5) التجارب العملية وتدريب الشبكات العصبونية، ونعرض في الفقرة (6) النتائج والمناقشة، وأخيراً نختم في الفقرة (7).

## 2- الدراسة المرجعية

اتجه الباحثون في البداية لطرق معالجة الإشارة التقليدية مثل تحليل الخصائص السمعية للإنسان Computational Auditory Scene Analysis (CASA) وخوارزميات التحليل إلى مصفوفات غير سالبة مثل Non-negative Matrix Factorization (NMF) والطرق الإحصائية مثل التحليل إلى المكونات المستقلة Independent Component Analysis (ICA). لم تفلح الطرق التقليدية في تقديم حل مرض للمسألة حيث كانت تفترض شروطاً لحل المسألة، كأن تتوفر معلومات مسبقة عن كل متكلم أو أن يتوفر أكثر من ميكرفون للتسجيل. فضلاً عن ذلك، فإن السمات التي تُستخرج بطرق معالجة الإشارة التقليدية مثل الطيف ومعاملات الميل كبستروم Mel Frequency Cepstrum Coefficients (MFCC) وغيرها، تتأثر بشكل كبير بالعديد من العوامل مثل الضجيج والعمر والحالة النفسية والصحية للمتكلم، مما دعا إلى ضرورة البحث عن سمات أعمق باستخدام شبكات عصبونية عميقة تُدعى تلك السمات الأشعة المضمنة Embedding vectors.

حقق الذكاء الصنعي قفزة كبيرة في هذا المجال منذ عام 2016 عن طريق استخدام نماذج ذكية تستخرج سمات مضمنة تميز كل متكلم عن الآخر. تنوعت الأبحاث بين طرق

تعتمد نماذج ذكية في المجال الترددي كما في [2, 3, 4, 5, 6, 7, 8] وأخرى تعتمد نماذج في المجال الزمني حيث تُرمّز الإشارة باستخدام نوع من الشبكات العصبونية هي المرمّزات encoder ثم تُستخرج السمات المضمنة كما في [9, 10, 11, 12, 13, 14]. كانت النتائج جيدة جداً في تلك الأعمال عندما تُختبر في نفس الظروف التي تدربت عليها وهي إشارات المزج التركيبية، حيث استطاع البعض الوصول إلى القيمة 3.51 للمعيار Perceptual Evaluation of Speech Quality (PESQ) الذي يعبّر عن مفهومية الكلام كما في [15]. تختلف إشارة المزيج التركيبية عن الحقيقية؛ فالأولى تُصمّم عن طريق جمع إشارات المتكلمين جمعاً رياضياً باستخدام برنامج محاكاة حاسوبي، والثانية تُسجّل باستخدام ميكروفون يلتقط إشارات المتكلمين ويخزنها في ملف حيث تتعرض الإشارات المسجّلة بالميكروفون لتشوهات لا خطية بسبب بيئة التسجيل والميكروفون المستخدم، كما تتعرض لتأخير زمني بسبب المسافة بين منبع الصوت والميكروفون. ينخفض أداء الشبكات العصبونية التي تتدرب على مدوّنات تدريب تركيبية كثيراً عند اختبارها في ظروف التسجيل الحقيقية [16].

حاول الباحثون بناء مدوّنات تدريب حقيقية تحتوي إشارات مزيج مسجّلة بميكروفون مع الأصوات المفردة الموافقة لكل مزيج. ولكن صعوبة الوصول إلى الأصوات المفردة حالت دون ذلك؛ إذ يجب على المتكلم أن يلفظ الجملة مرتين، مرة مع متكلم آخر لتسجيل إشارة المزيج وأخرى لوحده لتسجيل الصوت المفرد (الهدف)، ولكن من المستحيل أن يستطيع أحد ما، نطق نفس الجملة مرتين بشكل متطابق في المدة الزمنية و/أو المطال و/أو التردد. حاول بعض الباحثين بناء مدوّنات تشابه بيئات التسجيل الحقيقية من ناحية الضجيج فقط، فظهرت مدونات ضجيج مثل WHAM [17] و WHAMR [18] حيث سُجّلت إشارات ضجيج حقيقية من البيئات المحيطة، ثم أُستخدمت إشارات الضجيج السابقة لتشكيل

مدونات مضجّجة مثل LibriMix [19]. قرر آخرون تسجيل الإشارات الكلامية مع الضجيج مباشرة وليس جمعها رياضياً كما في LibriMix فظهرت المدوّنة CHiME-3 [20]، والمدوّنة CHiME-5 [21]، وMixer6 [22]، وVoxCeleb [23]. إنّ القيمة المضافة في المدوّنات الآنفة الذكر هي معالجة الضجيج، حيث تمّت إضافته للإشارات الكلامية المفردة إما جمعاً رياضياً باستخدام الحاسوب أو تسجيله مع الإشارات الكلامية المفردة مباشرة، ثم يجري جمع رياضي للإشارات الكلامية المفردة المضجّجة مع بعضها لتشكيل إشارات مزيج، أي أنها مدونات تركيبية مضجّجة وليست حقيقية، ودُربت الشبكات العصبونية في بعض الحالات على أهداف مع ضجيج كما في [24]، كانت الفائدة من المحاولات السابقة هي فصل الضجيج والقدرة على الفصل في بيئات صاخبة. إلا أنّ دقة فصل المتكلمين قد بقيت دون المطلوب.

بعد ذلك، فكر الباحثون بالاتجاه نحو التعلّم دون إشراف Unsupervised Learning وذلك للتعامل مع إشارات المزيج الحقيقية دون الحاجة لأصوات مفردة. قدّم الباحثون في [25] طريقة فصل متكلمين تعتمد على التعلم دون إشراف، حيث يتم التنبؤ بجنس كل متكلم أولاً باستخدام شبكة عصبونية، ثم يجري الفصل بناء على جنس المتكلم المستخرَج باستخدام شبكة عصبونية أخرى. في [26] أُستخدمت الشبكات التوليدية العدائية Generative Adversarial Networks إلى جانب التعلم دون إشراف واختير تابع خطأ مناسب ليقلل التشوهات الناتجة عن الفصل. اقترح الباحثون في [27] طريقة Mixture of Mixtures أي مزج مزيجين أو أكثر ثم إعادة فصلهم بطريقة تكرارية، أدى ذلك لظهور مشكلة جديدة هي الفصل الزائد over-separation أي الفصل لعدد أكثر من المتكلمين الموجودين فعلياً في المزيج. تصدّى الباحثون في [28] لمشكلة الفصل الزائد باستخدام النموذج معلّم–طالب Teacher-Student حيث دُرّب نموذج المعلّم في البداية ثم

استخدمت نتائج الفصل كأهداف تقريبية لنموذج الطالب. تمّ في [29] استخدام شبكتين مختلفتين في الخواص Heterogeneous Neural Networks لإنشاء أشباه أهداف pseudo labels من ملفات المزيج الحقيقية ثم تُحسّن تلك الأهداف بشكل تكراري. على الرغم من الجهود الحثيثة في استخدام طرق التعلم دون إشراف في فصل المتكلمين إلا أنها لم تعط حتى الآن نتائج أفضل من الطرق التي تعتمد على التعلم مع إشراف Supervised، علاوة على ذلك فالمسألة أصبحت أكثر صعوبة.

نطرح في هذا العمل طريقة هي الأولى حتى الآن لبناء مدوّنة تدريب حقيقية مع الأصوات المفردة، وذلك باستخدام تابع ضمن برنامج MATLAB الذي يتخاطب مع بطاقة الصوت ويستطيع تشغيل وتسجيل الملفات الصوتية بآن واحد وفي الزمن الحقيقي. قبل شرح ذلك التابع سنبين في الفقرة (3) المدوّنات الحقيقية والتركيبية والفرق بينهما.

## 3- المدوّنات التركيبية والمدوّنات الحقيقية

تحتوي مدوّنات التدريب، الخاصة بمسألة فصل المتكلمين، على ملفات المزيج وملفات المتكلمين (ملفات الهدف) ground truths. يتم تشكيل ملفات المزيج عن طريق جمع ملفات المتكلمين جمعاً رياضياً باستخدام برامج المحاكاة الحاسوبية كالماتلاب مثلاً، لذلك تسمى هذه المدونات بالتركيبية synthetic. تختلف ملفات المزيج التركيبية عن ملفات المزيج الواقعية (الحقيقية) أو التي تسمى أحياناً التلفيفية، حيث أن الميكروفون لا يجمع الإشارات الكلامية نقطة لنقطة بل تتعرض الإشارات لتشوهات عديدة خلال انتقالها من المتكلم وصولاً إلى الميكروفون. أضف إلى ذلك، أثر استجابة الميكروفون اللاخطية التي تتغير بتغيير الميكروفون. يمكن تقريب تلك التشوهات باستجابة نبضية لبيئة التسجيل حيث تكون إشارة المزيج هي مجموع جداءات التلاف للإشارات الكلامية

مع الاستجابة النبضية لبيئة التسجيل الموافقة لكل إشارة كما في المعادلة (1).

$$x(t) = \sum_{i=1}^{N} h_i(t) * s_i(t) + w(t) \qquad (1)$$

$x$ هي إشارة المزيج، $s_i$ إشارة المنبع رقم $i$. $w$ الضجيج. يشير الرمز(*) إلى عملية جداء التلاف convolution. $h_i$ الاستجابة النبضية لبيئة التسجيل الموافقة للمنبع $i$. $N$ عدد المتكلمين. المطلوب هو إيجاد $s_i$ من $x$.

يعتبر التعريف الرياضي السابق تقريباً للمزج الواقعي، على أية حال لا يوجد تعريف دقيق للمزج الواقعي وذلك لأنه يرتبط بعوامل عديدة منها: العوامل الجوية، الضجيج، المسافة بين المتكلم والميكروفون، عدد المتكلمين، حركة المتكلمين ...الخ. يمكن اعتباره تابع لا خطي ومتغير مع الزمن. جميع مدوّنات التدريب المستخدمة حتى الآن هي تركيبية ولكنها لا تعكس إشارات المزيج الحقيقية التي يسجلها الميكروفون لذلك ينخفض أداء أفضل نماذج فصل المتكلمين عند اختبارها على إشارات مزيج حقيقية (تلفيفية). كلما كانت بيانات مدوّنة التدريب أقرب للواقع كان أداء خوارزميات الفصل أدق في التجارب الواقعية.

## 4- الطريقة المقترحة لبناء مدوّنة حقيقية

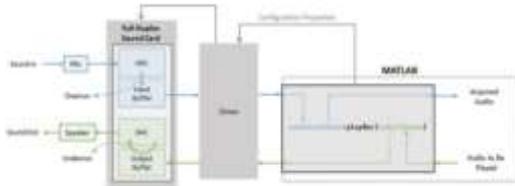

**الشكل(1) طريقة تخاطب برنامج الماتلاب مع بطاقة الصوت في الحاسب**

السبب الرئيسي في عدم وجود مدوّنة تدريب حقيقية هو عدم القدرة على إيجاد إشارات المتكلمين المفردة (الهدف) بشكل منفصل عند تسجيل إشارة المزيج، فالقيمة المضافة في هذه الورقة هي تسجيل وإيجاد تلك الإشارات المفردة. نبين في الفقرات التالية خطوات الطريقة المقترحة حيث نبدأ بشرح تابع التشغيل والتسجيل المستخدم، ثم نعرض شروط تسجيل المدوّنة لتوفير بيئة عمل مناسبة، بعد ذلك نبين طريقة اختيار الملفات

الصوتية اللازمة لبناء المدوّنة، وأخيراً استخلاص إشارات المتكلمين المفردة وتسجيل إشارة المزيج باستخدام تابع التشغيل والتسجيل.

## 4-1- تابع التشغيل والتسجيل

قمنا باقتراح طريقة لتسجيل إشارة مزيج مع الإشارات الهدف باستخدام تابع التشغيل والتسجيل المضمّن في برنامج الماتلاب AudioPlayerRecorder. يتميز هذا التابع بأنه يتخاطب مع بطاقة الصوت ثنائية الاتجاه Full duplex حيث يكتب ويقرأ عينات في/من بطاقة الصوت بنفس الوقت وفي الزمن الحقيقي. قدّمت شركة Mathworks في الإصدار رقم R2017a من برنامج MATLAB ذلك التابع الذي يشغّل ويسجل ملفات صوتية بنفس الوقت وبالزمن الحقيقي، حيث يتخاطب مع بطاقة صوت ثنائية الاتجاه Full duplex تحتوي على صوانين Buffer أحدهما لتخزين عينات الصوت القادمة من الميكرفون والآخر لتخزين عينات الصوت المراد إرسالها إلى جهاز إخراج الصوت كما يبين الشكل (1)

يتخاطب تابع الماتلاب AudioPlayerRecorder مع بطاقة الصوت بواسطة برنامج تعريف Driver حيث يحوّل تعليمات الماتلاب إلى لغة الآلة، توصي شركة Mathworks باستخدام برنامج التعريف /ASIO (Audio Stream Input Output) الذي يحقق أدنى تأخير زمني Latency للتسجيل.

يقوم التابع بقراءة ملف الصوت audioToDevice فيخزنه برنامج التعريف في صوان الخرج ثم يقوم المبدّل الرقمي التماثلي DAC (Digital/Analog Convertor) بتحويل الأطر المخزنة في الصوان إلى إشارات تماثلية يجري إرسالها إلى جهاز إخراج الصوت. يكون جهاز الميكرفون بنفس الوقت في حالة التقاط للصوت ثم يرسل الإشارة الملتقطة إلى المبدّل التماثلي الرقمي ADC (Analog/Digital Convertor) ومن ثم يجري تخزين عينات الصوت في صوان الدخل ويعيد برنامج التعريف ASIO الملف المسجّل إلى برنامج الماتلاب.

خلال العملية السابقة قد يحدث عدم تزامن في حالتين وهما:

امتلاء صوان الدخل Overrun: تحدث هذه الحالة إذا تأخر برنامج الماتلاب لسبب ما في قراءة صوان الدخل وما يزال الميكرفون في حالة تسجيل فقد يمتلئ صوان الدخل وعندها تضيع بعض من عينات الصوت.

خلو صوان الخرج Underrun: تحدث هذه الحالة عندما يتأخر برنامج الماتلاب في كتابة العينات في الصوان وعملية تشغيل الصوت جارية حينها قد يصبح الصوان فارغاً ويسجل الميكرفون لحظات صمت.

يزوّدنا تابع الماتلاب AudioPlayerRecorder بعدد العينات المفقودة في حالة Overrun وعدد عينات الصمت في حالة Underrun، فمن الممكن تدارك هاتين الحالتين.

إن الفكرة من استخدام التابع AudioPlayerRecorder هي القدرة على تشغيل ملف صوتي وتسجيله بالزمن الحقيقي ليكون هو ملف الهدف، بالإضافة إلى إمكان تشغيل ملفين معاً وتسجيل المزيج بملف يتضمن إشارة المزيج الحقيقي.

## 4-2- شروط تسجيل المدوّنة الحقيقية

يجب أن تكون عملية تسجيل مدوّنة التدريب الحقيقية دون انزياح زمني بين قناتي الخرج عند تسجيل إشارة المزيج ودون ظهور حالات Overrun و Underrun وذلك للحصول على مدوّنة بجودة عالية.

لكي نضمن أن التجربة تجري بالشكل المطلوب يجب أن نحقق الشروط التالية:

- وجود عتاد صلب يوفر سرعة معالجة مناسبة.
- نظام تشغيل مستقر.
- إيقاف جميع الإجرائيات والخدمات غير الضرورية التي تعمل على الحاسب لتخفيف حمل نظام التشغيل.
- التأكد أن تنجيز خوارزمية الماتلاب يعمل بالزمن الحقيقي عن طريق إجراء أمثلة لها.

لتنفيذ التجربة نحتاج إلى حاسوب بمواصفات عالية ومزوّد ببطاقة صوت لها مواصفات محددة (تردد تقطيع عالي وتدعم القنوات المتعددة multichannel).

استخدمنا حاسوباً يتمتع بالمواصفات التالية:

- اللوحة الأُم من طراز ROG STRIX Z390-F GAMING
- وحدة المعالجة المركزية CPU core i9
- حجم الذاكرة RAM 64GB
- بطاقة الصوت من طراز ROG SupremeFX 8-Channel High Definition Audio CODEC S1220A

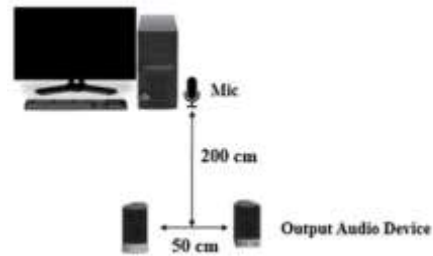

**الشكل (1) سيناريو التجربة لتسجيل ملفات المدوّنة الحقيقية**

تمّت عملية التسجيل بمخبر بعيد عن الضجيج وخالٍ من الأشخاص، وكانت المسافة بين الميكرفون وجهاز إخراج الصوت 2 متر والمسافة بين جهازي إخراج الصوت 50 سم كما هو موضح في الشكل (2).

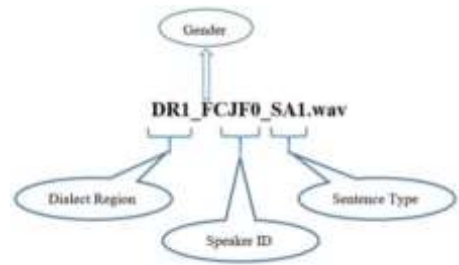

**الشكل (2) طريقة تسمية الملفات الصوتية لاستخدامها في بناء المدوّنة الحقيقية**

**4-3- اختيار الملفات الصوتية لبناء المدوّنة الحقيقية**

يمكن تطبيق الطريقة المقترحة لتصميم المدوّنة الحقيقية على أية ملفات صوتية من أي مدوّنة تدريب. استخدمنا مدوّنة

التدريب Texas Instruments/Massachusetts Institute of Technology (TIMIT) [30] حيث قمنا بما يلي:

1- إعادة تسمية الملفات الصوتية في المدوّنة بطريقة تُظهر هوية المتكلم وجنسه ولهجته والجملة المنطوقة كما في الشكل (3) وذلك لإنشاء قائمة بأسماء ملفات المزيج، علماً أن مدوّنة TIMIT تتضمن ثمانية لهجات أمريكية في مجلدات منفصلة.

2- إنشاء قائمة بأسماء ملفات المزيج التي سيتم تسجيلها بناءً على أربعة معايير مذكورة في [31] وهي:

- عدم تشكيل مزيج لنفس المتكلم ولجملتين مختلفتين.
- تنويع الجمل المحكية قدر المستطاع.
- تنويع المتكلمين قدر الإمكان في ملفات المزيج.
- اختيار ملفات صوتية متقاربة بالطول تجنباً لإضافة أصفار عند تشكيل المزيج.

**4-4- استخلاص ملفات المتكلمين الهدف**

إن الخطوة الأكثر أهمية في بناء المدوّنة الحقيقية هي استخلاص ملفات الهدف ground truths، فكل مدوّنة تدريب لفصل المتكلمين يجب أن تحتوي ملفات مزيج كدخل للنموذج العصبوني وملفات المتكلمين كخرج له. في حالة التسجيل الواقعي لملفات المزيج يصبح من الصعب استخلاص تسجيل كل متكلم لوحده. إن الطريقة المقترحة في هذا العمل تستطيع إيجاد ملفات الهدف باستخدام التابع AudioPlayerRecorder حيث يشغّل التابع ملف كل متكلم لوحده ويسجله باستخدام الميكرفون بملف جديد يكون هو الملف الهدف. ثم يعيد تشغيل الملفين معاً، كل ملف على قناة، وبنفس الظروف التي تم فيها تسجيل ملفات الهدف، ويسجّل الخرج فيكون هو ملف المزيج. إن ملفات الهدف هي ملفات صوتية واقعية، مسجلة بعد تشغيل ملفات المدوّنة TIMIT ضمن الشروط الواقعية. المسافة بين الميكرفون وجهاز إخراج الصوت هي 2 متر، وهذه الإشارات هي أكثر تشابها مع إشارات المزيج.

```
    oRun – overRun lost samples number.
    uRun – underRun lost samples number.
*/
6:  if (oRun OR uRun) > 0 then go to 5
7:  store gts1 in GTS
8:  gts2, oRun, uRun = APR(spk2)
    // gts2 – ground truth for speaker2.
9:  if (oRun OR uRun) > 0 then go to 8
10: store gts2 in GTS
11: rMix, oRun, uRun = APR(spk1 & spk2)
    // rMix – realistic Mixture.
12: if (oRun OR uRun) > 0 then go to 11
13: store rMix in RealMix
14: i ← i + 1
15:end while
```

أطلقنا اسم Realistic_TIMIT_2mix على المدوّنة الحقيقية الناتجة وهي تحتوي 30 ساعة تدريب training و10 ساعات اختبار validation و5 ساعات اختبار.

## 5– التجارب العملية

لاختبار تحسّن أداء نماذج فصل المتكلمين باستخدام المدوّنة الحقيقية قمنا بالخطوات التالية:

– بناء مدوّنة تركيبية Synthetic_TIMIT_2mix من نفس الملفات التي أُستخدمت لبناء المدوّنة الحقيقية بحيث كل ملف مزيج حقيقي يقابله ملف مزيج تركيبي من نفس ملفات المتكلمين.

– اختيار شبكة عصبونية لفصل المتكلمين وتدريب نسختين منها إحداها على المدوّنة الحقيقية والأخرى على التركيبية. اخترنا الشبكة العصبونية المقترحة في [32] كنموذج قاعدي للاختبار.

– المقارنة بين النموذجين المدرّبين من حيث دقة فصل المتكلمين باستخدام المعيار Scale Invariant Signal to Distortion Ratio (SI-SDR).

– المقارنة بين النموذجين من حيث استقرار الأداء مع تغير المسافة بين الميكرفون والمتكلمين.

## 4–5– تسجيل إشارة المزيج

يوفر تابع AudioPlayerRecorder وظيفة هامة وهي تشغيل ملفين صوتيين معاً، كلٌّ على قناة، ويتم تسجيل إشارة المزيج بنفس الوقت. لتنفيذ هذه التجربة يجب أن تدعم بطاقة الصوت تشغيل الصوت على قناتين Multichannel حيث يكون لكل قناة قسم خاص لها في صوان الخرج. يجب أن يتحقق التزامن بين القناة اليسرى واليمنى لوحدة إخراج الصوت خلال تشغيل الأصوات، وإلا فسيؤدي ذلك إلى انخفاض جودة ملفات الهدف عند تسجيلها.

لمعرفة الانزياح الزمني بين القناتين استخدمنا راسم إشارة لإظهار الإشارتين الصادرتين على القناتين معاً، (راسم الاهتزاز المستخدم هو من طراز BK Precision 2566-MSO بتردد تقطيع 2GS/s) ولم نلاحظ وجود أي انزياح بين القناتين. إن بطاقة الصوت المستخدمة المذكورة آنفاً ( ROG SupremeFX-8-Channel High Definition Audio CODEC S1220A ) تدعم تردد تقطيع حتى 192 KHz، وهو تردد عالٍ مقارنة بتردد تقطيع ملفات الصوت والذي هو 8KHz وهذا ما ساعد في ضبط التزامن بين القناتين.

تبين الخوارزمية 1 الطريقة المقترحة لبناء المدوّنة الحقيقية:

---

**Algorithm 1** *Creating Realistic_TIMIT_2mix*
**Input:**
- *S* – set of wav files from TIMIT train folder.
- *L* – mixtures names list.
**Output:**
- *GTS* – Folder containing ground truths for speakers.
- *RealMix* – Folder containing the realistic mixtures.

---

```
1: i ← 0
2: while i < length(L) do
3:   Extract wav files (spk1, spk2) from S correspond
     L(i).
4:   Down-sampling spk1 & spk2 to 8KHz
5:   gts1, oRun, uRun = APR(spk1)
     /* gts1 – ground truth for spk1.
```

نفصّل فيما يلي النموذج القاعدي المستخدم في فصل المتكلمين الذي سنعتمده للمقارنة بين المدونتين التركيبية والحقيقية، من خلال عمليات تدريب هذا النموذج.

### 5-1- النموذج القاعدي المستخدم لفصل المتكلمين

يتمثل النموذج بشبكة عصبونية عوديّة تتألف من طبقة دخل وأربع طبقات وحدات تكرارية ذات بوابات ثنائية الاتجاه Bidirectional Gated Recurrent Units (BGRU) وطبقة التوصيل الكامل Fully Connected Layer (FC). يبين الشكل (4) المخطط العام لعمل النموذج، حيث أن دخل الشبكة العصبونية هو مطال طيف إشارة المزيج بالمقياس اللوغاريتمي، وتمرّر بشكل سلاسل أشعة طول كل منها 129 ويُمرر كل شعاع على أربع طبقات BGRU، وبعدها يتم إسقاط شعاع الخرج في فضاء بُعده 20 (هو فضاء السمات المضمنة) باستخدام طبقة التوصيل الكامل، وعند الانتهاء من تمرير سلاسل طيف إشارة المزيج تُجرى عملية استخراج مراكز المتكلمين ضمن فضاء السمات المضمنة باستخدام الأقنعة الثنائية الموافقة لإشارة الدخل والتي صُممت خلال تجهيز معطيات التدريب (يعبّر القناع عن مصفوفة انتقائية تعطي عند ضربها بطيف المزيج طيف المتكلم الموافق). بعد إيجاد مراكز المتكلمين يتم تجميع أشعة السمات المضمنة حسب بُعدها عن تلك المراكز، حيث يتم بناء الأقنعة المقدَّرة Estimated Masks باستخدام المسافة بين المراكز وأشعة فضاء السمات. يُحسب الخطأ بين القناع الهدف والقناع المقدَّر، ثم بحساب مشتق تابع الخطأ خلال عملية الانتشار الخلفي Back Propagation تُعدّل أوزان النموذج بهدف تصغير الخطأ. إن حساب الخطأ بين القناع الهدف والقناع المقدَّر يكافئ حسابه بين الإشارة الهدف والإشارة المفصولة لأن القناع هو تمثيل للإشارة فيكفي ضرب القناع بمصفوفة الطيف وباستخدام تحويل فورييه العكسي نحصل على الإشارة الكلامية المفصولة، لذلك يمثل هذا النموذج نظام من طرف إلى طرف end-to-end.

تم اعتماد طبقات الشبكة العصبونية كما يلي:

❖ **طبقة الدخل:**

تم اعتماد لوغاريتم مطال طيف الإشارة الكلامية كدخل للشبكة العصبونية. يُحسب طيف الإشارة باستخدام تحويل فورييه القصير الأمد STFT من العلاقة التالية:

$$X_n(e^{jw_k}) = \sum_{m=-\infty}^{+\infty} w(n-m)x(m)e^{-jw_k m} \quad (2)$$

تردد التقطيع هو 8KHz النافذة المستخدمة هي Hanning وطولها 32ms أي 256 عينة.

نسبة التراكب 75% فيكون مقدار الإزاحة 8ms وتقابل 64 عينة.

يتم تطبيق تحويل فورييه القصير الأمد على المقطع الصوتي فيُمثَّل الخرج ببيان ثلاثي الأبعاد، البعد الأفقي هو الزمن والبعد العمودي يمثل التردد والبعد الثالث هو الطاقة ويُعبَّر عنها بتدرج الألوان فشدة اللون تدل على شدة الطاقة عند الترددات المحددة. لتكن أبعاد مصفوفة الطيف (m,n) فإن m هو عدد النوافذ الزمنية المتراكبة (طول المقطع الصوتي بالعينات مقسوماً على عدد عينات النافذة مضروباً بـ 4 بسبب التراكب 75%). طول المقطع الصوتي المستخدم 17460 عينة، فيكون في حالتنا عدد النوافذ $m = \left(\frac{17460}{256}\right) * 4 = 273$، أما n فهو عدد الترددات الموجودة في كل نافذة وهو في حالتنا 129 لأن عدد عينات النافذة الواحدة 256 فيكون عدد الترددات المفيدة مع المركبة الترددية عند الصفر DC هو 129=1+256/2.

❖ **طبقات BGRU:**

تم استخدام أربع طبقات BGRU ثنائية الاتجاه وذلك للاستفادة من معلومات الماضي والمستقبل في نمذجة السياق. تتألف كل طبقة GRU من 300 عصبون وبالتالي تتكون كل

طبقة BGRU من 600 عصبون، أي أن طول شعاع الحالة المخفية $h$ هو 600. إن طول شعاع دخل الطبقة الأولى 129 وخرجها 600، أما الطبقات الثانية والثالثة والرابعة فطول كل من شعاع الدخل والخرج هو 600 [32].

❖ **طبقة التوصيل الكامل Fully Connected Layer**

تقوم هذه الطبقة بإسقاط السمات التي ولّدتها طبقات BGRU الأربعة في فضاء بُعده 20 كل قيمة سلمية من شعاع الدخل الذي طوله 129 ستقابل شعاع طوله 20 أي طول شعاع خرج طبقة التوصيل الكامل FC Layer هو 2580=20*129. اختير طول شعاع السمات المضمنة 20 بناء على [2] حيث تمّ تجريب القيم {5,10,20,40,60} وكان أفضلها القيمة 20.

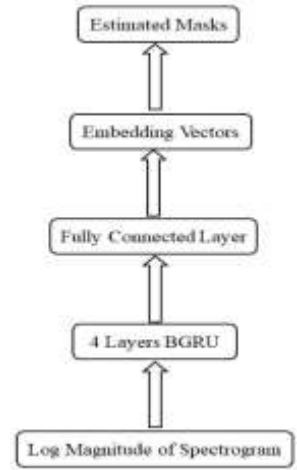

الشكل (3) النموذج العصبوني القاعدي المستخدم في التدريب

**5−2− تدريب النماذج**

دربنا نسختين من النموذج القاعدي، الأولى على المدوّنة التركيبية سنميه النموذج التركيبي Synthetic Model والنسخة الثانية على المدوّنة الحقيقية وسنسميه النموذج الحقيقي Realistic Model. معاملات النموذجين كانت متطابقة تماماً:

Epochs number =300, Optimizer Algorithm = ADAM [33], Batch size=128, Learning rate = $10^{-3}$

خلال التعلّم نغير قيمة معدل التعلّم للنصف عند مرور ثلاثة دورات دون أن يتناقص تابع خطأ التعميم validation error.

يبين الشكل (5) منحني التعلّم لكل من النموذجين التركيبي والحقيقي (يعطي قيمة تابع الفقد بدلالة عدد دورات التنفيذ). نلاحظ أن منحني النموذج التركيبي يصل إلى القيمة الصغرى (0.23) عند الدورة رقم 157، بينما يصل منحني النموذج الحقيقي إلى القيمة الصغرى (0.26) عند الدورة رقم 143. إنّ النموذج التركيبي كان أسرع في التقارب ولكن القيمة الصغرى لمنحني النموذج الحقيقي كانت هي الأقل. من الطبيعي أن يتقارب النموذج التركيبي قبل الحقيقي لأن المدوّنة التركيبية ناتجة عن محاكاة حاسوبية؛ فهي مثالية لا تحوي أي تشوهات بينما المدوّنة الحقيقية تمثل تسجيلات حقيقية بميكرفون وتحتوي بعض التشوهات من تخميد وتأخير زمني.

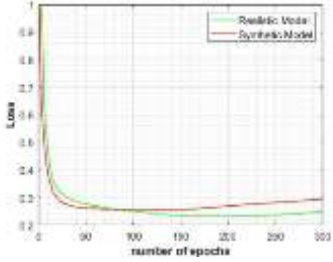

الشكل (4) منحني التعلم لكل من النموذج الحقيقي والتركيبي

**6− النتائج والمناقشة**

إن المعيار المستخدم في مسألة فصل المتكلمين هو نسبة الإشارة للتشوه بالمقياس الثابت Scale Invariant Signal to Distortion Ratio (SI-SDR) ويأخذ قيماً حقيقية، وكلما زادت قيمته كانت دقة الفصل أفضل [34]، ويُعرّف بالعلاقة (5):

$$s_{target} = \frac{\langle \hat{s}, s \rangle s}{\|s\|^2} \qquad (3)$$

$$e_{noise} = \hat{s} - s_{target} \qquad (4)$$

$$SI - SDR = 10 log_{10} \frac{\|s_{target}\|^2}{\|e_{noise}\|^2} \qquad (5)$$

تم الاختبار على نوعين من ملفات المزيج تركيبية وحقيقية. أُستخدمت المدوّنة LibriMix كملفات المزيج التركيبية، أما

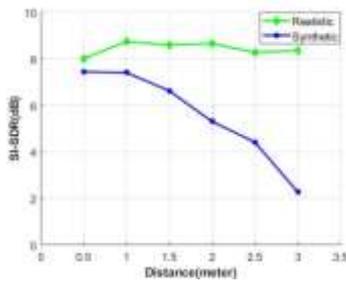

**الشكل (5) تغير أداء فصل المتكلمين مع تغير المسافة**

نلاحظ من الشكل (6) أن أداء النموذج الحقيقي مستقر مع تغير المسافة مقارنة مع النموذج التركيبي، حيث أن دقة الفصل بقيت فوق 8dB أمّا بالنسبة للنموذج التركيبي فقد انخفضت دقة الفصل كثيراً حتى قاربت 2dB. يمكن تفسير ذلك بأن النموذج الحقيقي الذي جرى تدريبه على ملفات مزيج حقيقية تفصل مسافة 2 متر بين الميكرفون والمتكلم؛ قد أُجبر على تعلّم خصائص الإشارة الكلامية بعد تعرضها للتشوهات بسبب بيئة التسجيل، وهذا الشيء الذي لم يتوفر لدى النموذج التركيبي مما أدى إلى انحدار أدائه مع ازدياد المسافة.

## 7– الخاتمة والآفاق المستقبلية

حققنا في هذه الورقة تحسيناً في أداء نظام ذكي لفصل المتكلمين عن طريق بناء مدوّنة تدريب حقيقية مع الأصوات المفردة، وهي الأولى من نوعها. حيث تمّ استخدام تابع من برنامج الماتلاب يقوم بكتابة وقراءة عينات الصوت في/من بطاقة الصوت بالزمن الحقيقي. قمنا باستغلال هذا التابع لطرح طريقة لتصميم أول مدوّنة حقيقية لمسألة فصل المتكلمين، وقد أدّى التدريب على تلك المدوّنة إلى تحسين دقة الفصل في حالة المزج الحقيقي بمقدار 1.65dB (حسب المعيار SI-SDR) مقارنة مع التدريب على مدوّنة تركيبية تقليدية. نرغب في المستقبل إجراء تعديلات على النموذج العصبوني المستخدم وإدراج الضجيج ضمن المدوّنة الحقيقية لمقاربة بيئات التسجيل الحقيقية الصاخبة.

الملفات الحقيقية فقمنا بتسجيلها بنفس الطريقة المذكورة في الخوارزمية 1 ولكن باستخدام ملفات صوتية لم تُستعمل للتدريب.

يبين الجدولان 1 و 2 نتائج هذه الاختبارات.

**الجدول (1) نتائج اختبار النموذجين التركيبي والحقيقي على المدوّنة LibriMix**

| النموذج | SI-SDR(dB) |
|---|---|
| التركيبي | 12.40 |
| الحقيقي | 13.65 |

**الجدول (2) نتائج اختبار النموذجين التركيبي والحقيقي على ملفات مزيج حقيقية**

| النموذج | SI-SDR(dB) |
|---|---|
| التركيبي | 7.01 |
| الحقيقي | 8.66 |

نلاحظ من الجدولين أن أداء النموذج الحقيقي أفضل من التركيبي في حالة المزج التركيبي والحقيقي، حيث تفوق النموذج الحقيقي على التركيبي بـ 1.65 dB حسب المعيار SI-SDR في حالة المزج الحقيقي وبمقدار 1.25dB في حالة المزج التركيبي وهي قيم تُعتبر هامة جداً في مسألة فصل المتكلمين. تثبت قيم الجدولين 1 و 2 فعّالية المدوّنة الحقيقية في تعليم النماذج العصبونية بشكل أفضل لفصل المتكلمين.

من جهة أخرى، نعلم أن المسافة بين المتكلم والميكرفون تؤثر كثيراً على دقة فصل المتكلمين، لذلك أجرينا اختبار النموذجين على مسافات مختلفة {0.5, 1, 1.5, 2, 2.5, 3} متر، ورسمنا المعيار SI-SDR بدلالة المسافة فحصلنا على الشكل (6)

# References

[1] E. C. Cherry, "Some Experiments on the Recognition of Speech, with One and with Two Ears,," *J. Acoust. Soc. Am.,,* vol. 25, no. 5, p. 975–979, Sep. 1953.

[2] J. R. Hershey, Z. Chen, J. L. Roux and S. Watanabe, "DEEP CLUSTERING:DISCRIMINATIVE EMBEDDINGS FOR SEGMENTATION AND SEPARATION," *IEEE International Conference on Acoustics, Speech and Signal Processing (ICASSP),* 2016.

[3] Z. Chen, Y. Luo and N. Mesgarani, "Deep attractor network for single-microphone speaker separation," in *In 2017 IEEE International Conference on Acoustics, Speech and Signal Processing (ICASSP) (pp. 246-250). IEEE,* 2017.

[4] C. Subakan, M. Ravanelli, S. Cornell, M. Bronzi and J. Zhong, "ATTENTION IS ALL YOU NEED IN SPEECH SEPARATION," *arXiv:2010.13154v1 ,* 25 Oct 2020.

[5] G.-P. Yang, C.-I. Tuan, H.-Y. Lee and L.-s. Lee, "Improved Speech Separation with Time-and-Frequency Cross-domain Joint Embedding and Clustering," *arXiv:1904.07845,* 16 Apr 2019.

[6] D. Yu, M. Kolbæk, T. Zheng-Hua and J. Jensen, "Permutation invariant training of deep models for speaker-independent multi-talker speech separation," *International Conference on Acoustics, Speech and Signal Processing (ICASSP),* 2017 IEEE .

[7] M. Kolbæk, D. Yu, Z.-H. Tan and J. Jensen, "Multitalker speech separation with utterance-level permutation invariant training of deep recurrent neural networks," *IEEE/ACM Transactions on Audio, Speech, and Language Processing ,* vol. 25, no. 10, October 2017).

[8] Z.-Q. Wang, S. Cornell, S. Choi, Y. Lee, B.-Y. Kim and S. Watanabe, "TF-GRIDNET: MAKING TIME-FREQUENCY DOMAIN MODELS GREAT AGAIN FOR MONAURAL SPEAKER SEPARATION," *arXiv:2209.03952,* 15 Mar 2023.

[9] Y. Luo and N. Mesgarani, "TASNET: TIME-DOMAIN AUDIO SEPARATION NETWORK FOR REAL-TIME, SINGLE-CHANNEL SPEECH SEPARATION," *IEEE, ICASSP 2018,* 2018.

[10] Y. Luo and N. Mesgarani, "Conv-TasNet: Surpassing Ideal Time–Frequency Magnitude Masking f or Speech Separation," *IEEE/ACM TRANSAC TIONS ON AUDIO, SPEECH, AND LANGUAGE P ROCESSING,* vol. 27, no. 8, AUGUST 2019.

[11] Y. Luo, Z. Chen and T. Yoshioka, "DUAL-PATH RNN: EFFICIENT LONG SEQUENCE MODELING FOR TIME-DOMAIN SINGLE-CHANNEL SPEECH SEPARATION," *IEEE, ICASSP 2020.*

[12] J. Chen, Q. Mao and D. Liu, "Dual-Path Transformer Network: Direct Context-Aware Modeling for End-to-End Monaural Speech Separation," *arXiv:2007.13975,* 14 Aug 2020.

[13] N. Zeghidour and D. Grangier, "Wavesplit: End-to-End Speech Separation by Speaker Clustering," *IEEE/ACM Transactions on Audio, Speech, and Language Processing 29,* pp. 2840-2849, 2021.

[14] J. Luo, J. Wang, N. Cheng, E. Xiao, X. Zhang and J. Xiao, "Tiny-Sepformer: A Tiny Time-Domain Transformer Network for Speech Separation," *arXiv:2206.13689,* 30 Jun 2022.

[15] Y. Liu and D. Wang, "Divide and Conquer: A Deep CASA Approach to Talker-independent Monaural Speaker Separation," *IEEE/ACM Transactions on Audio, Speech, and Language Processing,* 2019.

[16] C. Subakan, M. Ravanelli, S. Cornell and F. Grondin, "REAL-M: TOWARDS SPEECH SEPARATION ON REAL MIXTURES," *arXiv:2110.10812 ,* 20 Oct 2021.

[17] G. Wichern, J. Antognini, M. Flynn, L. Richard Zhu, E. McQuinn, D. Crow, E. Manilow and J. Le Roux, "WHAM!: Extending Speech Separation to Noisy Environments," *arXiv:1907.01160,* 2 Jul 2019.

[18] M. Maciejewski, G. Wichern, E. McQuinn and J. Le Roux, "WHAMR!: NOISY AND REVERBERANT SINGLE-CHANNEL SPEECH SEPARATION," *ICASSP 2020-2020 IEEE*

*International Conference on Acoustics, Speech and Signal Processing (ICASSP),* IEEE, 2020.

[19]  J. Cosentino, M. Pariente, S. Cornell, A. Deleforge and E. Vincent, "LibriMix: An Open-Source Dataset for Generalizable Speech Separation," *arXiv:2005.11262,* 22 May 2020.

[20]  J. Barker, R. Marxer, E. Vincent and S. Watanabe, "THE THIRD 'CHIME' SPEECH SEPARATION AND RECOGNITION CHALLENGE: DATASET, TASK AND BASELINES," *IEEE Workshop on Automatic Speech Recognition and Understanding (ASRU).,* 2015.

[21]  J. Barker, S. Watanabe, E. Vincent and J. Trmal, "The fifth 'CHiME' Speech Separation and Recognition Challenge: Dataset task and baselines," *arXiv preprint arXiv:1803.10609,* 2018.

[22]  L. Brandschain, D. Graff, C. Cieri, K. Walker, C. Caruso and A. Neely, "The Mixer 6 Corpus:Resources for Cross-Channel and Text Independent Speaker Recognition," *Proceedings of the Seventh International Conference on Language Resources and Evaluation (LREC'10).,* 2010..

[23]  A. Nagrani, J. Son Chung and A. Zisserman, "VoxCeleb: a large-scale speaker identification dataset," *arXiv:1706.08612,* 30 May 2018.

[24]  M. Maciejewski, J. Shi, S. Watanabe and S. Khudanpur, "TRAINING NOISY SINGLE-CHANNEL SPEECH SEPARATION WITH NOISY ORACLE SOURCES: A LARGE GAP AND A SMALL STEP," *arXiv:2010.12430,* 22 Feb 2021.

[25]  Y. Wang, J. Du, L.-R. Dai and C.-H. Lee, "A Gender Mixture Detection Approach to Unsupervised Single-Channel Speech Separation Based on Deep Neural Networks," *IEEE/ACM Transactions on Audio, Speech, and Language Processing ,* vol. 25, no. 7, July 2017.

[26]  K. Saijo and T. Ogawa, "REMIX-CYCLE-CONSISTENT LEARNING ON ADVERSARIALLY LEARNED SEPARATOR FOR ACCURATE AND STABLE UNSUPERVISED SPEECH SEPARATION," *arXiv:2203.14080,* 26 Mar 2022.

[27]  S. Wisdom, E. Tzinis, H. Erdogan, R. J. Weiss, K. Wilson and J. R. Hershey, "Unsupervised Sound Separation Using Mixture Invariant Training," *arXiv:2006.12701v2,* 24 Oct 2020.

[28]  J. Zhang, C. Zorila, R. Doddipatla and J. Barker, "Teacher-Student MixIT for Unsupervised and Semi-supervised Speech Separation," *arXiv:2106.07843,* 9 Sep 2021.

[29]  J. Han and Y. Long, "Heterogeneous separation consistency training for adaptation of unsupervised speech separation," *EURASIP Journal on Audio, Speech, and Music Processing,* no. 1, pp. 1-17, 2023.

[30]  J. S. Garofolo, L. F. Lamel, W. M. Fisher, J. G. Fiscus, D. S. Pallett and N. L. Dahlgren, "DARPA TIMIT acoustic-phonetic continous speech corpus CD-ROM. NIST speech disc 1-1.1," *NASA STI/Recon technical report n, 93, 27403,* 1993.

[31]  M. Maciejewski, G. Sell, L. P. Garcia-Perera, S. Watanabe and S. Khudanpur, "BUILDING CORPORA FOR SINGLE-CHANNEL SPEECH SEPARATION ACROSS MULTIPLE DOMAINS," *arXiv:1811.02641,* 6 Nov 2018.

[32]  R. Melhem, A. Jafar and R. Hamadeh, "Improving deep attractor network by BGRU and GMM for speech separation," *Journal of Harbin Institute of Technology (New Series),* vol. 28, no. 3, pp. 90-96, 2021.

[33]  D. P. Kingma and J. Ba, "Adam: A Method for Stochastic Optimization," *arXiv:1412.6980v9,* 30 Jan 2017.

[34]  J. Le Roux, S. Wisdom, H. Erdogan and J. R. Hershey, "SDR – HALF-BAKED OR WELL DONE?," *ICASSP,* IEEE 2019.